\documentclass[prd,twocolumn,superscriptaddress,longbibliography,nofootinbib,notitlepage]{revtex4-1}

\usepackage{amsmath,amssymb,mathrsfs}
\usepackage[euler]{textgreek}
\usepackage{psfrag}
\usepackage{graphicx}
\usepackage{color}
\usepackage[usenames,dvipsnames]{xcolor}
\usepackage{pifont}
\usepackage{fourier}
\usepackage{xr}
\usepackage{stmaryrd}
\usepackage{cancel}

\usepackage{fontenc}
\usepackage{graphicx}
\usepackage{bbding}
\usepackage{lettrine}
\usepackage[a-1b]{pdfx}
\usepackage{setspace}
\usepackage{upgreek}
\usepackage{amsfonts}
\usepackage{braket}
\usepackage{quoting}
\usepackage[normalem]{ulem}
\usepackage[multiple]{footmisc}
\usepackage{mathtools}
\usepackage{amsmath}
\usepackage[margin=1in]{geometry}
\usepackage{amsmath,amssymb,amsfonts}
\usepackage{listings}
\usepackage{xcolor}
\usepackage{booktabs}
\usepackage{siunitx}
\usepackage{float}  
\usepackage{tikz}
\usepackage{verbatim}
\usetikzlibrary{arrows,shapes,backgrounds}
\usepackage{tikzpagenodes} 
\usepackage{fancyhdr}
\usepackage[process=auto]{pstool}
\usepackage[normalem]{ulem}

\usepackage{times}
\usepackage{euscript,array}
\usepackage{bm}

\definecolor{codegreen}{rgb}{0,0.6,0}
\definecolor{codegray}{rgb}{0.5,0.5,0.5}
\definecolor{codepurple}{rgb}{0.58,0,0.82}
\definecolor{backcolour}{rgb}{0.95,0.95,0.92}
\definecolor{blu}{rgb}{0,0,0.75}

\lstdefinestyle{fortranstyle}{
    backgroundcolor=\color{backcolour},
    commentstyle=\color{codegreen},
    keywordstyle=\color{magenta},
    numberstyle=\tiny\color{codegray},
    stringstyle=\color{codepurple},
    basicstyle=\ttfamily\footnotesize,
    breakatwhitespace=false,
    breaklines=true,
    captionpos=b,
    keepspaces=true,
    numbers=left,
    numbersep=5pt,
    showspaces=false,
    showstringspaces=false,
    showtabs=false,
    tabsize=2,
    language=Fortran
}

\newcommand{\bea}{\begin{eqnarray}}
\newcommand{\eea}{\end{eqnarray}}
\newcommand{\beq}{\begin{equation}}
\newcommand{\eeq}{\end{equation}}\usepackage{graphicx}
\usepackage{dcolumn}
\usepackage{bm,morefloats,color}
\newcommand{\g}{\mathcal{G}}
\newcommand{\mr}{\mathcal{R}}

\newcommand{\ea}[1]{\begin{eqnarray}#1\end{eqnarray}}
\newcommand{\eq}[1]{\begin{equation}#1\end{equation}}

\def\X5sp{{\rm X}_5}
\def\Y3sp{{\rm Y}_3}
\def\Z3sp{{\rm Z}_3}

\usepackage{hyperref}

\newcommand{\chg}[1]{#1}

\begin{document}

\title{Scalar \chg{vacuum polarization in loop-quantum-gravity black holes}}

\author{Antonino Flachi}
\email{flachi@keio.jp}
\affiliation{Department of Physics, \chg{and} Research and Education Center for Natural Sciences, Keio University, 4-1-1 Hiyoshi, Yokohama, Kanagawa 223-8521, Japan}
\author{Marco Pasini}
\email{pasini.marco@spes.uniud.it}
\affiliation{Department of Mathematics, Computer Science, and Physics, University of Udine, Via delle Scienze 206, I-33100 Udine, Italy}

\date{\today}

\begin{abstract}
A quantum macroscopic ``Kruskal'' black hole solution that incorporates quantum geometry effects has been derived in \chg{loop quantum gravity} as the counterpart to the classical Schwarzschild solution with a distinct imprint outside the event horizon, even at scales much larger than the Planck length. This resulting black hole quantum geometry is supported by an effective energy density of quantum origin, outside the horizon, which prevents asymptotic flatness at large distances and confines massive particles to finite radii, thereby preventing their escape to infinity. In this work\chg{,} we adapt to these solutions the extended Anderson-Candelas-Christensen-DeWitt approach to compute the quantum vacuum polarization in
order to provide an accurate measure of the quantum activity around these black holes. 
We carry out a numerical implementation of the formalism and present, to our knowledge for the first time, the scalar vacuum polarization $\langle\phi^2\rangle$ exterior to this quantum-corrected geometry. We find that the quantum-gravity exponent $\epsilon$ enhances the near-horizon polarization and induces, farther out, a small negative tail that we identify---through a parameter-free \chg{DeWitt-Schwinger} comparison---with the field's response to the nonzero curvature of the background (absent for Ricci-flat Schwarzschild). The correction scales linearly with $\epsilon$, the parameter tracking the quantum gravitational corrections, so that\chg{,} for astrophysically realistic (i.e., tiny) $\epsilon$, the result is numerically indistinguishable from Schwarzschild. The calculation furnishes a consistency check on the quantum activity around these solutions, the fluctuations tracking the local curvature without anomalous growth in the exterior.

\end{abstract}

\maketitle

\section{Introduction}
\label{sec:introduction}
\chg{Loop quantum gravity} is an example of a theory that, in principle, enables self-consistent calculations of quantum gravitational effects, free from \textit{ad hoc} phenomenological parametrizations or educated guesses, by assuming a fundamentally discrete spacetime structure. This capability is a particularly valuable feature to inspect quantum aspects of black holes, since quantum gravitational corrections can leave detectable imprints at macroscopic scales well beyond the Planck length, outside the horizon in regions that can\chg{, in principle,} be subject to observation. Thus, scrutiny of quantum phenomena in the exterior regions of quantum-corrected black holes is especially warranted, as certain proposals (\cite{LQG1, LQG2, LQG3, LQG4, Suddho}) reveal that interior quantum corrections can trigger anomalously large fluctuations in low-curvature regions and offer a basis for refining or discarding the underlying theoretical framework.

A new compelling example is the black hole solution of Refs.~\cite{AO1,AO2,AO3,FaraoniGiusti}, which replaces the classical Schwarzschild metric with a quantum-corrected counterpart explicitly designed to eliminate the central singularity while avoiding the anomalously large quantum corrections in low-curvature regions. This example is  noteworthy for offering a glimpse into quantum processes near strong gravitational sources. Concretely, the \chg{quantum-gravity} corrections of this solution are measured by a small dimensionless parameter $\epsilon$\chg{:}
$$
1 + \epsilon = \sqrt{1+ \left( \frac{{\gamma^2 \Delta}}{\chg{2\pi M^2}}\right)^{1/3}},
$$ 
where the Newton constant and the speed of light have been set to unity, $M$ is the black hole mass, $\gamma \simeq 0.2375$ \chg{is} the Barbero-Immirzi parameter\chg{,} and the constant $\Delta \approx 5.17 \ell_{Pl}^2$ is the minimum positive eigenvalue of the area operator with $\ell_{Pl}$ being the Planck length. Treating $\Delta$ as small, i.e., at large scales, we have at leading order $\epsilon \approx \sqrt[3]{\left( { \gamma^2 \Delta}/{16\pi M^2} \right)}$, and, correspondingly, 
the quantum-corrected black hole metric is approximated by
the following static, spherically symmetric line element\chg{:}
\bea
ds^2= r^{2\epsilon} f(r)d\tau^2 + r^{-2} u(r) dr^2 + r^2 d\Omega_2^2,
\label{metric2}
\eea
where the unit radius \chg{two-sphere} line element has been indicated with $d\Omega_2^2 = d\theta^2+\sin(\theta)^2d\phi^2$ and 
\bea
\label{f}
f(r) &=& r_s^{-2\epsilon} \left[ 1- \left(\frac{r_s}{r}\right)^{1+\epsilon}  \right] \\
\label{u}
u(r) &=& {r^{2}\over r_s^{2\epsilon}} f^{-1}(r).
\eea
with $\epsilon$ a positive real number. 

The location of the unique event horizon of the above quantum black hole solution coincides with the Schwarzschild one: $r_s=2 M >0$. 
For large astronomical black holes, $\epsilon$ is tiny due to the $M^{-2/3}$ dependence. Differently from the classical Schwarzschild behavior, there is here an effective negative quantum energy density outside the horizon given by $\rho = - \left({\epsilon}/{8\pi r^2} \right) \left({r_s}/{r} \right)^{1+\epsilon}$,
that sustains the solution, preventing asymptotic flatness at large distances. Relevant analysis has been carried out in Ref. \cite{FaraoniGiusti}, where it was confirmed that this negative purely quantum-mechanical energy density yields \chg{timelike} radial geodesics with massive particles experiencing a repulsion from the accumulated negative energy, confining them to finite radii and preventing escape to infinity (null geodesics, i.e.\chg{,} photons, propagate outward but arrive at infinity with infinite redshift). Further quantitative analyses of Ref. \cite{FaraoniGiusti} revealed a
\chg{quasilocal} mass that vanishes at infinity, signaling a loss of asymptotic flatness and persistent quantum effects in low-curvature regions. This does not \textit{necessarily} rule out these solutions or\chg{, more generally,} loop quantum gravity, but it does make a weak field analysis a rather subtle process. 

A key question concerns how quantum effects behave in the presence of such black holes. The goal of this work is to investigate in detail the quantum vacuum polarization effects exterior to the aforementioned black hole solution, thereby providing deeper insight into the quantum activity around these quantum gravitational black holes. While computing vacuum polarization effects in curved spacetimes poses a significant technical challenge\chg{,} in general, established methods for evaluating quantum effects near black holes have been developed over the past several decades. Here, we adopt the approach pioneered by Candelas~\cite{Candelas:1980}, who devised a systematic framework that integrates the covariant point-separation regularization scheme of DeWitt and Christensen~\cite{DeWitt:1965d,Christensen} with the WKB approximation. Candelas' seminal study focused on Schwarzschild black holes and subsequent refinements, chiefly those of Anderson et al. \cite{Anderson:1990jh}, extended Candelas' framework, overcoming some of its original limitations and broadened its applicability to a variety of black hole geometries. We refer to this approach as the extended Anderson-Candelas-Christensen-DeWitt formalism (extended CCDeW formalism, for short). A review of the existing literature is outside the scope of this paper\chg{; here,} we limit ourselves to citing a few examples.
{Applications and refinements of the (extended) CCDeW mode-sum formalism include Refs.~\cite{Anderson:1990jh,Shiraishi,Tomimatsu:2000,Winstanley:2008sr,Flachi:2008sr,Cvetic:2015,Quinta:2016,Flachi:2015,Morley:2021}, whereas methodologically distinct approaches to the vacuum polarization, based on alternative mode-sum or \chg{Green-Liouville} prescriptions, are developed\chg{, for instance,} in Refs.~\cite{Frolov,Breen:2016,Breen,Thompson}.} To the best of our knowledge, the vacuum polarization has not previously been computed for the loop-quantum-gravity geometry~(\ref{metric2}): while $\langle\phi^2\rangle$ has been obtained for asymptotically Lifshitz black holes~\cite{Quinta:2016} and, recently, for the regular Bardeen geometry~\cite{BoassoMazzitelli}, and quantum fields on other loop-quantum-gravity black holes have been examined at the level of scattering cross sections and Hawking emission~\cite{Moulin:2019}, the metric structure~(\ref{metric2}) differs from all of these and has not been discussed. Here, we follow closely the methodology outlined in Ref.~\cite{Flachi:2008sr}\chg{,} to which we refer the reader for full details. Schematically, one starts from a sum-over-modes representation of the Green function and extracts the coincidence limit by incorporating \chg{counterterms} that cancel the associated divergences. These \chg{counterterms}, computed for general four-dimensional backgrounds~\cite{Christensen}, yield a regularized result upon subtraction, which remains challenging to evaluate due to divergences in the mode summation. These are addressed by approximating the summand via WKB techniques and by isolating the divergent contributions that are handled through analytical continuation. The remainder, regular by construction, is then computed numerically: this is the computationally intensive step that entails solving a large number of differential equations with singular coefficients and performing summations over slowly converging series. While optimizations are possible, such as the \chg{Monte Carlo--inspired} techniques used in Ref.~\cite{Flachi:2008sr}, here we adopt a ``lower level'' \chg{high-performance} implementation of the calculation. The formalism we adopt and derivations for the present case trace closely Ref.~\cite{Flachi:2008sr} and are only briefly detailed in Sec.~\ref{sec2}. In Sec.~\ref{sec3}, we describe the numerical implementation and present the results. Finally, in Sec.~\ref{sec4}\chg{,} we draw our conclusions. 
{}\\
\section{\label{sec2} Extended Anderson-Candelas-Christensen-DeWitt Formalism}

In this section we describe the formalism, originally developed in Ref.~\cite{Candelas:1980} for Schwarzschild black holes and later refined in Ref.~\cite{Anderson:1990jh} for asymptotically flat geometries and in Ref.~\cite{Flachi:2008sr} for the case of asymptotically \chg{anti--de Sitter} black holes. In this section, we fix the notation and describe the main steps of the calculations. The underlying spacetime metric we consider is of the form (\ref{metric2}). We make the assumption that there is at least one horizon, i.e.\chg{,} a real solution to the equation $f(r_s)=0$, which is the case for $f$ given by (\ref{f}). 
Notice that in (\ref{metric2}) a rotation to imaginary time (i.e., to Euclidean space) $t \to -i \tau$ has been performed; this allows for the identification of an inverse temperature $\beta=1/T_H$ expressed as 
\bea
T_H = \beta^{-1} &=& {1\over 4 \pi} {g_{tt,r} \over \sqrt{g_{tt} g_{rr}}}\Big|_{r=r_s} = {1+ \epsilon \over 4 \pi r_s},
\eea
which reduces to the usual Hawking temperature in the Schwarzschild limit $\epsilon \to 0$. 

The quantum state considered throughout is the \chg{Hartle-Hawking} state, which we define as follows. The exterior region of (\ref{metric2}) is static, with Killing vector $\partial_t$ and a \chg{nondegenerate} Killing horizon at $r=r_s$ of surface gravity $\kappa=(1+\epsilon)/(2r_s)$. The Euclidean section is regular at $r=r_s$ (i.e., free of a conical singularity) if and only if $\tau$ is periodic with period $\beta=2\pi/\kappa$. The \chg{Hartle-Hawking} state is the state whose two-point function is the analytic continuation of the unique Green function on this Euclidean section that is regular at the horizon and decays as $r\to\infty$; in practice, this amounts to the Matsubara decomposition with frequencies $\omega_n=2\pi n/\beta$ and to selecting radial modes that are regular at $r=r_s$ and decay at large $r$, as done below. The state so defined is invariant under the isometries of the background and thermal Kubo-Martin-Schwinger (KMS) at temperature $T_H=\kappa/2\pi$ with respect to $\partial_t$; on spacetimes with a bifurcate Killing horizon\chg{,} it is the unique regular (Hadamard) state with these properties~\cite{KayWald}, and its existence has been established rigorously in the static case~\cite{Sanders,Gerard}. Two remarks are in order. First, although the geometry is not asymptotically flat, $g_{tt}$ grows as $r^{2\epsilon}$, so that the local (Tolman) temperature $T_H/\sqrt{g_{tt}}$ redshifts to zero at large $r$ and, for a massive field, the decaying solution is selected unambiguously. Second, the definition \chg{involves only} the exterior effective geometry (\ref{metric2}), treated as a fixed background.

The problem we shall address in this work is the computation of the coincidence limit of the scalar thermal (i.e., Euclidean) Green's function\chg{:}
\bea
\langle \phi^2 \rangle
= \g\left(X,X\right) = \lim_{X' \to X} \g\left(X,X'\right)
\eea
with $\g\left(X,X'\right)$ obeying the following equation\chg{:}
\bea
\label{Eq:3}
\left(\Box
-m^2
-\xi \mr 
\right)\g\left(X,X'\right)=-{\det g}^{-1/2} \delta\left(X-X' \right),
\eea
where $X=(\tau,r,\theta,\phi)$, $\mr$ the Ricci scalar, which for the present case is given by
\bea
\mr = -\frac{2\epsilon}{r^{2}} \left[ \epsilon + 1 + \left( \frac{r_{s}}{r} \right)^{\epsilon + 1} \right]
\eea
and $\xi$ a \chg{nonminimal} coupling, which takes real values. Imposing spherical symmetry allows us to express the thermal Green's function $\g(X,X')$ as follows, by separation of variables:
\begin{equation}
\g(X,X')=\frac{1}{\beta}\sum_{n=-\infty}^\infty {e^{i\omega_n (\tau-\tau')}} \sum_{l=0}^\infty \frac{2l+1}{4 \pi} P_l(\cos \gamma )\g_{n,l}(r,r'), 
\label{Eq:9}
\end{equation}
where $\omega_n={2\pi n}/{\beta}{}\equiv n\alpha$ [with $\alpha\equiv 2\pi/\beta=(1+\epsilon)/(2r_s)$] and $n \in \mathbf{Z}$; $P_l$ represents a Legendre polynomial and  $\cos(\gamma)=\cos(\theta)\cos(\theta')+ \sin(\theta)\sin(\theta')\cos(\phi-\phi')$.
Using (\ref{Eq:9}) in (\ref{Eq:3}) and writing $\delta (\tau-\tau')=\frac{1}{\beta}\sum_{n=-\infty}^\infty e^{i\omega_n (\tau-\tau')}$ leads to the radial equation for $\g_{n,l}(r,r')$:
\begin{widetext}
\bea
&& \left\{\frac{d^2}{dr^2}
+
\left({1 \over r} + {1+\epsilon\over r} { 1 \over 1 - (r_s/r)^{1+\epsilon} }
\right) 
\frac{d}{dr}
-
\mathscr{U}
\right\}
\g_{n,l}(r,r') = - {1\over r^2} {(r_s/r)^{\epsilon} \over 1 - (r_s/r)^{1+\epsilon} } {\delta(r-r')},\label{Eq:5}
\eea
where
\ea{
\mathscr{U} = \left(
\left({r_s\over r }\right)^{2 \epsilon} \frac{\omega_n^2} { \left( 1 - (r_s/r)^{1+\epsilon}\right)^2}
+ { m^2+ l(l+1) r^{-2} \over 1 - (r_s/r)^{1+\epsilon} }
- \frac{2\xi\epsilon}{r^{2}} \left[ \epsilon + 1 + \left( \frac{r_{s}}{r} \right)^{\epsilon + 1} \right]
{1 \over 1 -  (r_s/r)^{1+\epsilon}}
\right).
}
\end{widetext}

The solution $\g_{n,l}(r,r')$ to Eq.~(\ref{Eq:5}) can be expressed in terms of the solutions 
to the associated homogeneous equation,
that can be classified by the respective asymptotic behavior far away from the black hole and at the horizon.
Here, we shall label $p_{nl}(r)$ the solution regular at the event horizon (and diverging at infinity) and $q_{nl}(r)$ the solution diverging at the horizon (and regular at infinity). These solutions are independent. From Eqs.~(\ref{Eq:9}) and (\ref{Eq:5})\chg{,} we can write the full Green function as follows\chg{:}
\begin{equation}
\label{Eq:10}
\g(X,X') = \frac{1}{\beta}
\sum_{n=-\infty}^\infty 
e^{i\omega_n (\tau-\tau')} 
\sum_{l=0}^\infty 
\frac{2l+1}{4 \pi} 
P_l(\cos(\gamma))
\g_{n,l}(r,r'),
\end{equation}
where the radial Green's function takes the form:
\begin{equation}
\label{Eq:19}
\g_{n,l}(r,r') = \frac{1}{r^2}
\frac{\left({r_s/r}\right)^\epsilon }{1-(r_s/r)^{1+\epsilon}}
\frac{p_{nl}(r_<) q_{nl}(r_>)}{q_{nl} p'_{nl}- p_{nl}q'_{nl} },
\end{equation}
where $r_>$ and $r_<$ denote the larger and the smaller of the pair $\{r,r'\}$, respectively.

The coincidence limit that can be calculated starting from (\ref{Eq:10}) requires a knowledge of the behavior of the solutions asymptotically and at the horizon, which need to be implemented also in the numerical calculation. 

The asymptotic, large-distance behavior of the solutions $\phi_{nl}^{(\infty)}(r)$ is determined at leading order by
\ea{
\left\{\frac{d^2}{dr^2}
+
{2+\epsilon\over r} \frac{d}{dr}
-\left(m^2 
+ \left( {r_s \over r} \right)^{2\epsilon} \omega_n^2
\right)
\right\}\phi_{nl}^{(\infty)}(r) = 0,
\label{Eq:14}
}
where terms of \chg{the order of} $\text{O}\left(r^{-(1+\epsilon)}\right)$ have been ignored. The coefficient of the first derivative with respect to $r$ in (\ref{Eq:14}) is universal as long as the behavior of the function $f(r)$, given by (\ref{f}) in the present case, is analytic and yields higher order corrections O$(r^{-2})$. We should notice that the above expansion is valid for any $\epsilon \geq 0$, but we have in mind the case of $\epsilon \ll 1$. 

The method we are using is valid\chg{,} in general, independently of the asymptotic behavior of the black hole solution, but its numerical implementation requires imposing at infinity the correct behavior of the solution, which depends on the geometry. While the implementation of the asymptotic behavior is\chg{, in general,} a simple procedure \textit{per se}, the complications arise since the computation of the vacuum polarization requires the contribution from large values of $l$ and $n$. This is a delicate point even for the relatively simple solution (\ref{metric2}) we are considering here. For the case of $\epsilon=0$, which reduces the problem to Schwarzschild, the correct asymptotic behavior, ignoring terms of \chg{the order of} $l(l+1)r^{-2}$, is $\phi_{nl}^{(\infty)} \sim \exp \left( \pm \sqrt{m^2+\omega_n^2}\,r\right)/\sqrt{r^2(m^2+\omega_n^2)}$. Since the computation requires summing up contributions to the vacuum polarization up to large values of $l$, i.e.\chg{,} solving the mode equation up to large $l$, this implies that the outer boundary has to be chosen accordingly so to make the approximation of the asymptotic behavior consistent. With the parameter $\epsilon \neq 0$ switched on, the \chg{subleading} term to the Schwarzschild asymptotic acquires a dependence on $r$, i.e.\chg{,} becomes of \chg{the order of} $\omega_n^2 r^{-2\epsilon}$. This means that the leading asymptotic behavior for small $\epsilon$ and large $r$ remains the same as Schwarzschild, but the \chg{subleading} terms are of \chg{the order of} $\epsilon \log r$. This again requires the numerical implementation of the boundary behavior at large distance $r_{asy}$ from the horizon to be imposed in accordance with the condition $\epsilon \omega_n^2 \log {r_{asy}} \ll 1$. 

Imposing numerically the regularity of the solutions at the horizon is a much easier task. We can expand in the near-horizon region and directly solve the equation. This is done by first introducing the following ``tortoise''-type coordinate:
\bea
f_* = r^{\epsilon} {f},
\eea
and
\bea
dr_* = {dr \over f_*}
\eea
which allows one to express the radial equation as follows\chg{:}
\bea
&&\left\{\frac{d^2}{dr_*^2}
- r_s^{-2\epsilon} \left(r^{2\epsilon} f\right)
\left(
{l(l+1) \over r^2}
+ r_s^{2\epsilon}  {df_* \over dr} {r^{-1-\epsilon}} 
+ m^2 + \xi \mr
\right) \right. \nonumber \\
&&\left. - r_s^{-2\epsilon}  \omega_n^2
\right\}\left(
r \phi(r) \right) = 0. \nonumber
\eea
In the near-horizon limit the existence of a horizon implies that $f$ asymptotes to zero\chg{,} and in the present case we also notice that $f \times u$ asymptotes, in the same near-horizon limit, to a constant. This gives at leading order for $r \to r_s$:
\bea
\left\{\frac{d^2}{dr_*^2}
- r_s^{-2\epsilon}\omega_n^2
\right\}\left(
r \phi^{(hor)}_{nl}(r) \right) = 0.
\eea
Thus\chg{, the leading-order} asymptotic near-horizon solution is
\bea
\phi^{(hor)}_{nl}(r) \sim {e^{\pm \omega_n r_* / r_s^{\epsilon}}\over r}.
\eea

The problem with directly evaluating (\ref{Eq:10}) numerically is that it is divergent and requires regularization. One possibility is to find some approximate solutions and manage the diverging behavior explicitly. This step of the procedure is not unique, partly because of the \chg{nonuniqueness} of regularization\chg{, in general,} and partly because any approximation scheme that allows one to manage the subtraction of the divergences would be acceptable. Here, we adopt the WKB form of the normal modes that provides a sufficiently general framework to handle the problem more or less directly at least in four dimensions. We\chg{, thus,} express the general solution of the homogeneous equation associated with (\ref{Eq:5}) as follows: 
\begin{equation}
\begin{split}
\label{soluz}
\phi(r)=r^{A}W(r)^{B} \exp\left(\pm\int_{r_s}^r W(r')^C r'^{D} u(r')^E f(r')^L\, dr'\right)
\end{split}
\end{equation}
where the $\pm$ signs refer to the solutions regular or singular at the horizon, respectively. 
Inserting (\ref{soluz}) in the homogeneous equation and fixing the coefficients by requiring that mixed signature terms vanish \chg{[}this is uniquely achieved by setting $A = -(\epsilon+2)/2$, $B = -1/2$, $C = 1$, $D=-1$, $E=1/2$, \chg{and} $L=-1/2$\chg{]} yields an equation for $W$:
\eq{
W^2 =  \varpi + \sigma + a_1 {W'\over W} +a_2 {W^{'2}\over W^2} +a_3 {W^{''}\over W}~,
\label{wkbeqt}
}
where 
{\ea{
\varpi &=& \left[\left(l+{1\over 2}\right)^2-{1\over 4}\right] {f \over r^2}+{\omega_n^2\over r^{2\epsilon}}~,\nonumber\\
\sigma &=& 
\left(m^2 + \xi \mr \right) {f}  +\left(1 + {\epsilon\over 2}\right)^2 {f\over u}
+ {r\over 2}\left(1+ {\epsilon \over 2}\right) {f'\over u} \nonumber\\
&&
- {r \over 2}\left(1+{\epsilon \over 2}\right) {f u'\over u^2}
~,
\nonumber}}
and
{\ea{
a_1 &=& {r_s^{2\epsilon}\over 2} f f',\nonumber\\
a_2 &=& -{3 r_s^{2\epsilon}\over 4} {f^2},\nonumber\\
a_3 &=& {r_s^{2\epsilon}\over 2}{f^2}.\nonumber
}}
\chg{Equation}~(\ref{wkbeqt}) is just a rewriting of the homogeneous equation associated to (\ref{Eq:5})\chg{,} and no approximation has yet been implemented. A solution can be found iteratively by writing 
\bea
W &=& W^{(0)} + W^{(1)} + \chg{\cdots}.
\eea
In order to extract the divergent contributions\chg{,} it is sufficient in four dimensions to include up to the \chg{next-to-leading-order} contribution leading to
\ea{
\tilde W^{-1}_{nl} &=& {1\over \sqrt{\Phi}} - {\Psi\over 4\sqrt{\Phi^3}}, 
}
where
\ea{
\Phi &=& \varpi + \sigma, \\
\Psi &=& a_1 {\Phi'\over \Phi} +\left({a_2-a_3 \over 2}\right) \left({\Phi'\over \Phi}\right)^2 + a_3 {\Phi''\over \Phi}.
}
We have used the notation $\tilde W_{nl}$ to signify that the WKB expansion is truncated: $\tilde W_{nl}$ is the ap\-proxi\-mate \chg{next-to-leading-order} solution to (\ref{wkbeqt}). Notice that this approximation is compensated in the expression for the coincidence limit of the Green's function since the WKB terms are added and subtracted: the calculation of the coincidence limit is \textit{exact}.

The renormalized coincidence limit of the Green's function can be computed using the general \chg{counterterms} of Ref.~\cite{Christensen} and following closely the implementation of the CCDeW as outlined in Ref.~\cite{Flachi:2008sr}. The explicit calculation has been discussed\chg{, in general,} in Ref.~\cite{Flachi:2008sr}. The calculation is lengthy but straightforward and yields the result: 
\ea{
\langle \phi^2 \rangle
=
{\alpha \over 8\pi^2}\left[
\Upsilon_0 + \Sigma_1 + \Sigma_2 - \Delta +\Theta
\right]\chg{,}
\label{renormalizedphisquare}
}
where 
\ea{
\Upsilon_0 \equiv \sum^{\infty}_{l=0}\left({l+1/2\over r^{\epsilon+2} \tilde{W}_{0l}}-{1 \over r^{\epsilon+1} \sqrt{f}}\right)~,
}
and 
\ea{
\Sigma_1
=
\sum_{l=0}^{\infty}
\left(l+{1\over 2}\right)
\left(
{1\over r^{\epsilon+2}W_{0l}} - {1\over r^{\epsilon+2}\tilde W_{0l}}
\right),
}
\ea{
\Sigma_2
=
2 \sum_{n=1}^{\infty}\sum_{l=0}^{\infty}
\left(l+{1\over 2}\right)
\left(
{1\over r^{\epsilon+2} W_{nl}} - {1\over r^{\epsilon+2}\tilde W_{nl}}
\right),
}
where the tilde indicates the WKB counterpart. The analogous quantity without the tilde represents the exact counterpart and is evaluated numerically. 
The expressions for the $\Delta$ and $\Theta$ terms are lengthy and are reported in the Appendix.
The interested reader can check, term by term, the correctness of the above results by comparing with previous results. Notice that all terms in (\ref{renormalizedphisquare}) are convergent and regular by construction.

\section{\label{sec3} Numerical analysis}

{We evaluate the renormalized coincidence limit~(\ref{renormalizedphisquare}) with a dedicated high-performance implementation. 
The numerically demanding ingredients are the exact radial modes entering $\Sigma_1$ and $\Sigma_2$: for each Matsubara frequency $\omega_n$ and angular number $l$\chg{,} we integrate the homogeneous radial equation associated with~(\ref{Eq:5}) by a fourth-order \chg{Runge-Kutta} scheme, starting from the regular near-horizon expansion in the tortoise coordinate and imposing at the outer boundary the asymptotic behavior dictated by~(\ref{Eq:14}); the next-to-leading WKB counterpart $\tilde W_{nl}$ is subtracted mode by mode. The \chg{counterterms} $\Delta$ and the analytic block $\Theta=P_1+P_2+P_3$ are computed from the closed-form expressions given in Sec.~\ref{sec2} and in the Appendix, the generalized $\zeta$ \chg{functions}~(\ref{Zdef}) being evaluated through their rapidly convergent Bessel representation. Throughout\chg{,} we report the dimensionless combination $\mathrm{vp}\equiv 16\pi^2 M^2\langle\phi^2\rangle$ as a function of $r/r_s$, and quote the field mass through $\mu\equiv mM=m/2$ (we work in units $r_s=2M=1$).}

{Because $\langle\phi^2\rangle$ arises as a \chg{near cancellation} between the (large) mode sum and the (large) covariant counterterms, the result carries an intrinsic relative noise floor, set by the consistency of the add-and-subtract procedure. Two points require care. First, the angular ($l$) sum in $\Sigma_2$ must be truncated at the grid-independent plateau of its cumulative value: beyond it, the high-$l$ radial modes accumulate \chg{Runge-Kutta} discretization error that, for $\epsilon\neq0$, contaminates the tail; we verified on three grids ($N=10^4,2\times10^4,4\times10^4$) that the plateau value is \chg{grid stable} to $<0.3\%$. Second, we confirmed convergence in the far zone: at $\epsilon=0.1$ the exterior profile is invariant to within $1.2\%$ under doubling the grid (\chg{$N=10^4\to2\times10^4$}) and to within $0.1\%$ under doubling the outer radius (\chg{$r_{\rm max}=200\to400r_s$}) throughout $r/r_s\in[3,10]$, so the results are quantitatively reliable well beyond the near-horizon region. As a validation, in the Schwarzschild limit $\epsilon\to0$ our horizon value $\mathrm{vp}\simeq1.05\times10^{-2}$ (at $\mu=1/2$) reproduces Anderson's result~\cite{Anderson:1990jh} to about $1\%$, the profile peaks at the horizon, and the large-$r$ tail agrees with the \chg{DeWitt-Schwinger} large-mass benchmark (see below). We have also reproduced all of Anderson's horizon values of $\mathrm{vp}$ given in \cite{Anderson:1990jh}.}

\begin{figure}[t]
\centering
\includegraphics[width=\columnwidth]{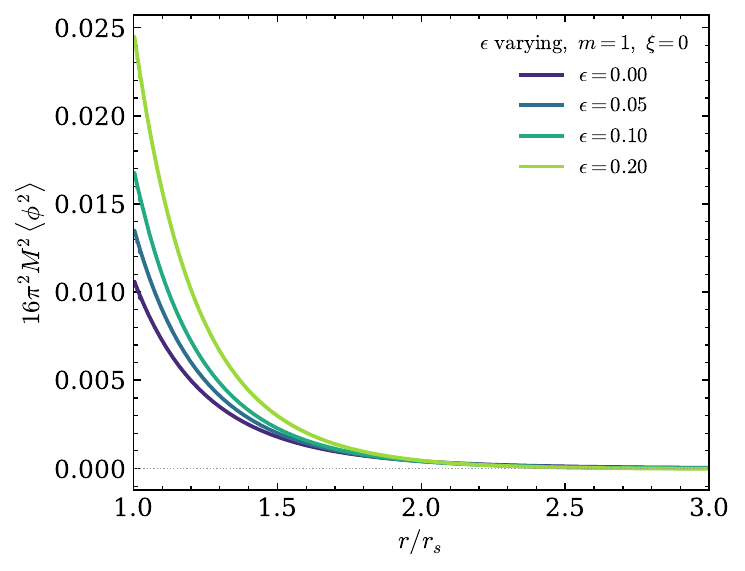}
\caption{{Vacuum polarization $16\pi^2M^2\langle\phi^2\rangle$ versus $r/r_s$ for $\epsilon=0,0.05,0.10,0.20$ at fixed $\mu=1/2$ ($m=1$ in the legend), $\xi=0$. The quantum-gravity exponent enhances the near-horizon polarization; $\epsilon=0$ is the Schwarzschild limit.}}
\label{fig:z}
\end{figure}

\begin{figure}[t]
\centering
\includegraphics[width=\columnwidth]{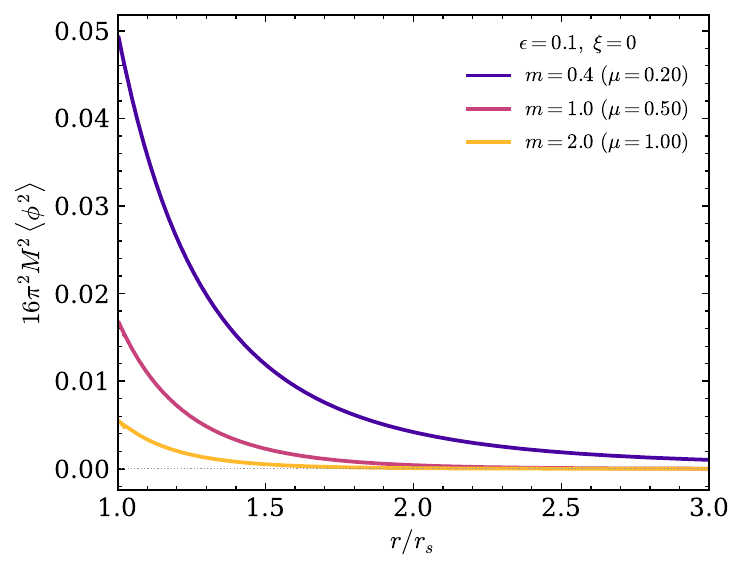}
\caption{{Dependence on the field mass, $\mu=0.2,0.5,1.0$, at $\epsilon=0.1$, $\xi=0$. Heavier fields are increasingly suppressed.}}
\label{fig:m}
\end{figure}

\begin{figure}[t]
\centering
\includegraphics[width=\columnwidth]{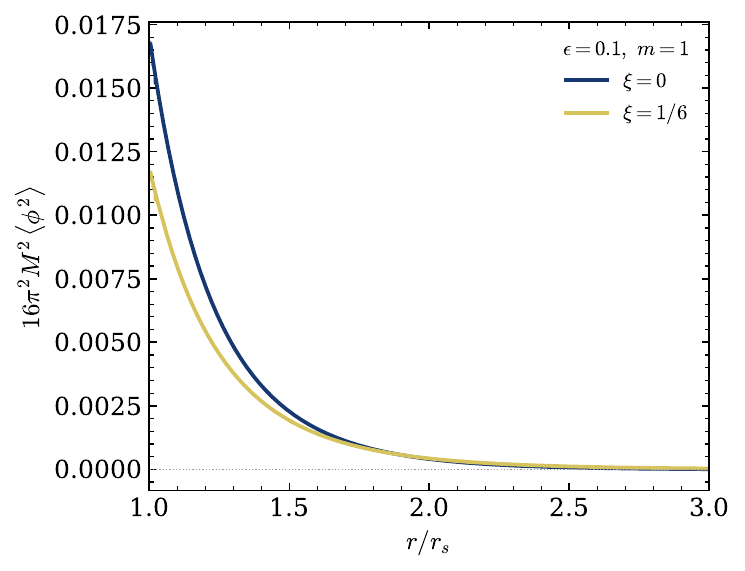}
\caption{{Effect of the curvature coupling, minimal ($\xi=0$) versus conformal ($\xi=1/6$), at $\epsilon=0.1$, $\mu=1/2$ ($m=1$ in the legend). Conformal coupling lowers the polarization near the horizon.}}
\label{fig:xi}
\end{figure}

\begin{figure}[t]
\centering
\includegraphics[width=\columnwidth]{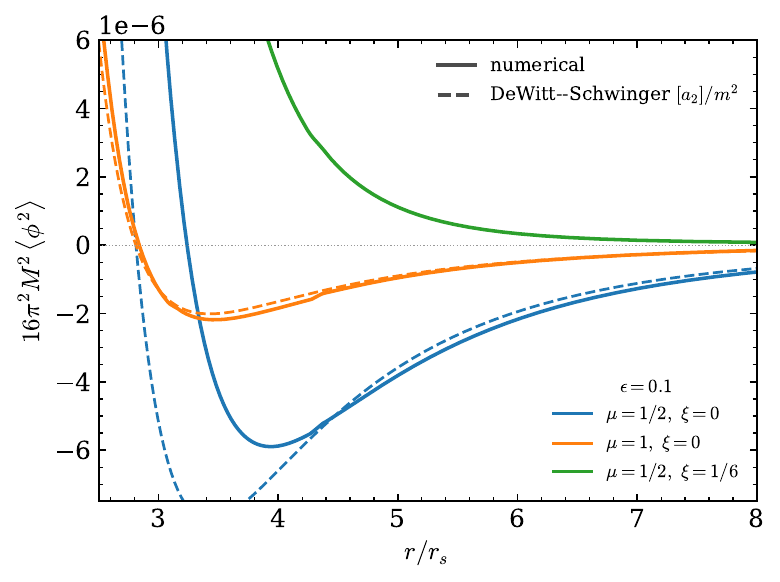}
\caption{{Large-$r$ behavior at $\epsilon=0.1$: for \chg{nonconformal} coupling\chg{,} the polarization turns \emph{negative} beyond $r/r_s\approx3$. \chg{Solid line:} full numerical $16\pi^2M^2\langle\phi^2\rangle$; \chg{dashed line:} the scheme-independent \chg{DeWitt-Schwinger} prediction $(1/4)[a_2]/m^2$ (no free parameters). The two agree in sign and to \chg{$\sim10\%$}, best for $\mu=1$. Conformal coupling $\xi=1/6$ (\chg{green line}) stays positive in the range shown: there\chg{,} the $[a_2]$ prediction is strongly suppressed and turns negative only for $r/r_s\gtrsim6.8$, at the $10^{-8}$ level, which is not resolved on this scale. The effect is the field's curvature response to the negative Ricci scalar of the \chg{loop-quantum-gravity} background, absent for Ricci-flat Schwarzschild.}}
\label{fig:negDS}
\end{figure}

{Figures~\ref{fig:z}--\ref{fig:xi} display the vacuum polarization for representative choices of the quantum-gravity exponent $\epsilon$, the field mass $\mu$, and the coupling $\xi$; Table~\ref{tab:vp} collects the corresponding values at four radii. Three robust trends emerge. (i) The loop-quantum-gravity exponent $\epsilon$ \emph{enhances} the near-horizon polarization: at the horizon $\mathrm{vp}$ grows by a factor \chg{of} $\sim2.3$ from $\epsilon=0$ to $\epsilon=0.2$ (Fig.~\ref{fig:z}). (ii) A heavier field \emph{suppresses} $\langle\phi^2\rangle$, the expected decoupling of massive fluctuations (Fig.~\ref{fig:m}). (iii) Conformal coupling $\xi=1/6$ \emph{lowers} the polarization relative to the minimal case (Fig.~\ref{fig:xi}). In every case the profile is finite, peaking at the horizon and decaying monotonically through the near zone. For $\epsilon\neq0$ with \chg{nonconformal} coupling, however, the decay does not terminate at zero: beyond $r/r_s\approx3$ (the precise location depends on the field mass: at $\epsilon=0.1$, $r/r_s\approx3.2$ for $\mu=1/2$ and $r/r_s\approx2.8$ for $\mu=1$)\chg{,} the polarization changes sign into a small \emph{negative} tail (Fig.~\ref{fig:negDS}), which we analyze below.}

\begin{table}[t]
\centering
\caption{{Scalar vacuum polarization $\mathrm{vp}=16\pi^2 M^2\langle\phi^2\rangle$ for the metric~(\ref{metric2}) at four areal radii $r/r_s$, for the parameter sets \chg{in} Figs.~\ref{fig:z}--\ref{fig:xi} ($\mu=mM$). The $\epsilon=0$ row is the Schwarzschild reference. Values are converged to the $\lesssim0.1\%$ noise floor in $r/r_s\in[1,2]$.}}
\label{tab:vp}
\begin{ruledtabular}
\begin{tabular}{ccc cccc}
$\epsilon$ & $\mu$ & $\xi$ & $1.005$ & $1.1$ & $1.5$ & $2.0$ \\
\hline
0.00 & 0.5 & 0   & \chg{$1.05\times10^{-2}$} & \chg{$7.24\times10^{-3}$} & \chg{$1.79\times10^{-3}$} & \chg{$4.13\times10^{-4}$} \\
0.05 & 0.5 & 0   & \chg{$1.34\times10^{-2}$} & \chg{$8.98\times10^{-3}$} & \chg{$2.01\times10^{-3}$} & \chg{$4.05\times10^{-4}$} \\
0.10 & 0.5 & 0   & \chg{$1.67\times10^{-2}$} & \chg{$1.10\times10^{-2}$} & \chg{$2.28\times10^{-3}$} & \chg{$4.06\times10^{-4}$} \\
0.20 & 0.5 & 0   & \chg{$2.43\times10^{-2}$} & \chg{$1.57\times10^{-2}$} & \chg{$2.98\times10^{-3}$} & \chg{$4.50\times10^{-4}$} \\
0.10 & 0.2 & 0   & \chg{$4.91\times10^{-2}$} & \chg{$3.58\times10^{-2}$} & \chg{$1.19\times10^{-2}$} & \chg{$4.18\times10^{-3}$} \\
0.10 & 1.0 & 0   & \chg{$5.41\times10^{-3}$} & \chg{$3.35\times10^{-3}$} & \chg{$5.12\times10^{-4}$} & \chg{$6.54\times10^{-5}$} \\
0.10 & 0.5 & 1/6 & \chg{$1.16\times10^{-2}$} & \chg{$7.95\times10^{-3}$} & \chg{$1.93\times10^{-3}$} & \chg{$4.29\times10^{-4}$} \\
\end{tabular}
\end{ruledtabular}
\end{table}

\begin{figure}[t]
\centering
\includegraphics[width=\columnwidth]{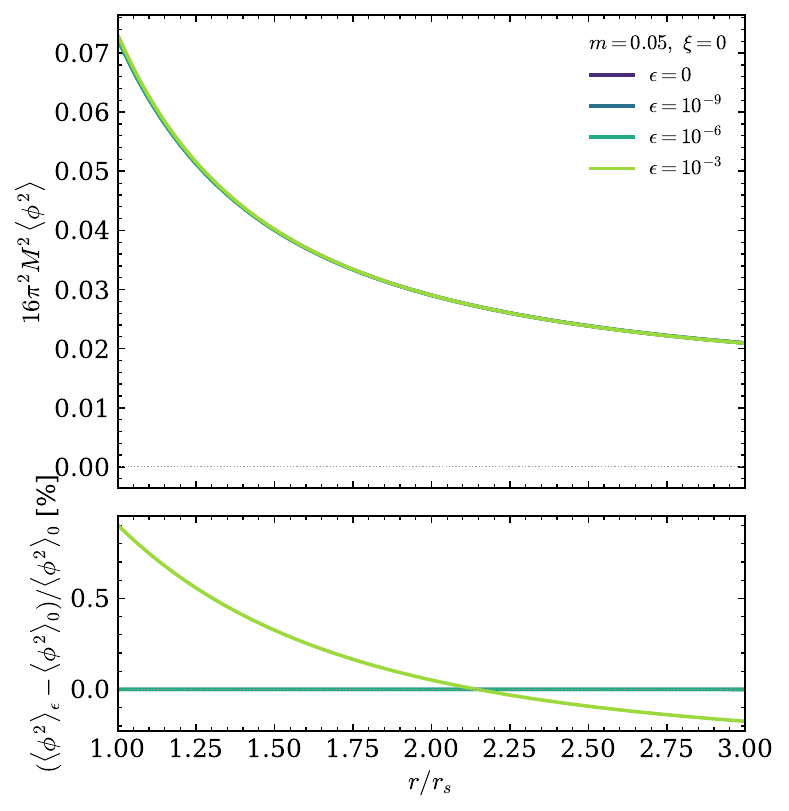}
\caption{{Light field ($\mu=0.025$, i.e.\chg{,} $m=0.05$ in the legend, $\xi=0$) at tiny $\epsilon=0,10^{-9},10^{-6},10^{-3}$. Top: the profiles coincide on the scale of the plot. Bottom: relative deviation from Schwarzschild, $(\langle\phi^2\rangle_\epsilon-\langle\phi^2\rangle_0)/\langle\phi^2\rangle_0$, showing the linear-in-$\epsilon$ correction.}}
\label{fig:light}
\end{figure}

{It is instructive to ask how these results behave in the physically relevant regime $\epsilon\ll1$. For an astrophysical black hole\chg{,} the $M^{-2/3}$ scaling makes $\epsilon$ exceedingly small (e.g.\chg{,} $\epsilon\sim10^{-26}$ for a solar mass), so one expects the polarization to be indistinguishable from the Schwarzschild value. We confirm this quantitatively in Fig.~\ref{fig:light}: for a light field, $\mu=0.025$, the relative deviation from the $\epsilon=0$ profile grows linearly with $\epsilon$, reaching only $\sim0.9\%$ at $\epsilon=10^{-3}$ and falling below $10^{-3}\%$ for $\epsilon\le10^{-6}$, i.e.\chg{,} below our noise floor. The leading quantum-gravity correction to $\langle\phi^2\rangle$ is\chg{, therefore,} $\mathcal{O}(\epsilon)$.}

{The sign change noted above is physically significant. Unlike Schwarzschild---\chg{Ricci flat}, with $\langle\phi^2\rangle$ positive throughout the exterior---the background~(\ref{metric2}) carries a nonzero \emph{negative} Ricci scalar, $\mr=-2\epsilon[\epsilon+1+(r_s/r)^{1+\epsilon}]/r^2$, which decays only as $r^{-2}$ and reflects the persistent negative quantum energy density of the solution~\cite{FaraoniGiusti}. The massive field responds through the curvature-coupling term of its renormalized vacuum polarization: the large-$r$ tail is reproduced by the scheme-independent part of the \chg{DeWitt-Schwinger} expansion~\cite{Christensen,AHS}, $\langle\phi^2\rangle_{\rm DS}\simeq[a_2]/(16\pi^2 m^2)$, with the heat-kernel coefficient $[a_2]=\tfrac12(\xi-\tfrac16)^2\mr^2+\tfrac16(\tfrac15-\xi)\Box\mr+\tfrac1{180}(R_{\mu\nu\rho\sigma}R^{\mu\nu\rho\sigma}-R_{\mu\nu}R^{\mu\nu})$. Evaluated for the metric~(\ref{metric2}), this predicts a negative tail (dominated by the $\Box\mr<0$ term) that agrees with the full numerical result in sign and to \chg{$\sim10\%$} in magnitude for $r/r_s\gtrsim4$ (Fig.~\ref{fig:negDS}), the agreement being best for the heavier field $\mu=1$ where the large-mass expansion is most accurate (ratio $\approx1.0$ over $r/r_s\in[3,7]$). Consistently, the effect is absent for Schwarzschild ($\mr=0$) and is strongly suppressed at conformal coupling $\xi=1/6$, where the $\mr^2$ term vanishes and the coefficient of $\Box\mr$ is reduced from $1/30$ to $1/180$: for $\epsilon=0.1$ the \chg{DeWitt-Schwinger} prediction then remains positive up to $r/r_s\approx6.8$ and turns negative only farther out, at the level $|\mathrm{vp}|\lesssim3\times10^{-8}$ (for $\mu=1/2$), too small to be resolved on the scale of Fig.~\ref{fig:negDS}. We also verified that the negative tail is invariant under doubling the grid and the outer radius. The negative region is\chg{, therefore,} a genuine imprint of the non-Ricci-flat quantum geometry, not a numerical artifact. For astrophysically realistic $\epsilon$\chg{,} both the near-horizon enhancement and this far tail are negligibly small (Fig.~\ref{fig:light}): the field fluctuations track the local curvature and decay, showing no anomalous growth in the accessible exterior despite the background's loss of asymptotic flatness---a clean consistency check on the quantum activity around these solutions.}

\section{\label{sec4} Conclusions}

{We have computed the scalar vacuum polarization $\langle\phi^2\rangle$ of a massive, \chg{nonminimally} coupled field in the \chg{Hartle-Hawking} state exterior to the loop-quantum-gravity black hole of Refs.~\cite{AO1,AO2,AO3,FaraoniGiusti}, adapting the extended \chg{Anderson-Candelas-Christensen-DeWitt} formalism to the metric~(\ref{metric2}) and implementing it as a high-performance numerical code. To the best of our knowledge this is the first such calculation for this geometry. The quantum-gravity exponent $\epsilon$ enhances the near-horizon polarization, a heavier field suppresses it, and conformal coupling reduces it; the profile is finite and peaks at the horizon. Farther out, for \chg{nonconformal} coupling, $\langle\phi^2\rangle$ develops a small \emph{negative} tail that we identify---through an independent, parameter-free \chg{DeWitt-Schwinger} comparison---with the field's curvature response to the nonzero negative Ricci scalar of the background; it is absent for Ricci-flat Schwarzschild and strongly suppressed at conformal coupling. The leading correction relative to Schwarzschild is linear in $\epsilon$, so that for astrophysically realistic values the polarization is indistinguishable from the classical result---a reassuring consistency check, given that the underlying geometry is neither \chg{Ricci flat} nor asymptotically flat: the quantum fluctuations track the local curvature and decay, without anomalous growth in the exterior. A natural continuation is the renormalized stress tensor $\langle T_{\mu\nu}\rangle$, which controls \chg{backreaction}; the generality of the present implementation makes its extension to other quantum-corrected and regular black holes straightforward. More generally, the method is not tied to the specific form~(\ref{metric2}) and can be applied to other metrics, including other loop-quantum-gravity-inspired black holes such as those of Refs.~\cite{ZhangLMY,HusainKSW}, provided that the geometry (or at least the exterior region of interest) is static and spherically symmetric. The presence and the nature of a horizon are not essential to the method: they \chg{change only} the boundary conditions imposed on the radial modes, so that the only model-dependent inputs are the metric functions and these boundary conditions.}

\section*{Acknowledgments}
This work was conducted using the Miyabi Supercomputing System at the Joint Center for Advanced High Performance Computing (JCAHPC). We are grateful to Stefano Ansoldi for discussions at the initial stages of this project and to Gon\c{c}alo Quinta for sharing an early Mathematica notebook.

\section*{\chg{Data Availability}}
\chg{The data are publicly available~\cite{zenodo}. The code is available upon request to the authors.}

\begin{widetext}
\appendix
\section*{Appendix: $\Delta$ and $\Theta$ terms} 

The $\Delta$ term is written as the sum of several contributions as follows:
\ea{
\Delta = \Delta_{11} + \Delta_{12} + \Delta_{21} + \Delta_{22} + \Delta_{31} + \Delta_{32} + \Delta_4}
with 
\ea{
\Delta_{11} \equiv -\sum^{\infty}_{n=1}\left[ {2\over r^{2\epsilon}f}\left(\sqrt{\omega^2_n+m^2r^{2\epsilon}f} -\omega_n-{m^2r^{2\epsilon}f \over 2\omega_n}\right) \right],
\nonumber
}

\ea{
\Delta_{12} \equiv -\sum^{\infty}_{n=1}\left[ \left(\xi-{1\over 6}\right)\mr\left({1\over \sqrt{\omega^2_n+m^2r^{2\epsilon}f}}-{1\over \omega_n}\right) \right],
\nonumber
}

\ea{
\Delta_{21} &\equiv& {m^2 \over 2\alpha} \ln\left(m^2r^{2\epsilon}f\right) - {m^2 \over \alpha} \ln\left(\alpha + (\alpha^2+m^2r^{2\epsilon}f)^{1/2}\right), \nonumber
}

\ea{
\Delta_{22} &\equiv& 
{2i\over r^{2\epsilon}f} \int^{\infty}_{0}{dt \over e^{2\pi t}-1}\left(
\left[(1+it)^2\alpha^2+m^2r^{2\epsilon}f\right]^{1/2}-\left[(1-it)^2\alpha^2+m^2r^{2\epsilon}f\right]^{1/2}\right),\nonumber
}

{\ea{
\Delta_{31} &\equiv&{1\over 2} {\left(\xi-{1\over 6}\right)\mr} \bigg[{\ln\left(m^2 r^{2\epsilon}f\right)\over \alpha}-{2\over \alpha}\ln\left(\alpha + (\alpha^2+m^2r^{2\epsilon}f)^{1/2}\right) +{1\over \sqrt{\alpha^2+m^2r^{2\epsilon}f}} \bigg],\nonumber
}}
\ea{
\Delta_{32} &\equiv& {\left(\xi-{1\over 6}\right)\mr} \bigg[
2\alpha i \int^{\infty}_{0}{dt \over e^{2\pi t}-1}\left({1\over \left[ (1+it)^2\alpha^2+m^2r^{2\epsilon}f\right]^{1/2}}-{1\over \left[(1-it)^2\alpha^2+m^2r^{2\epsilon}f\right]^{1/2}}\right)\bigg],\nonumber
}
{\ea{
\Delta_{4} &\equiv& 
{1\over 2\alpha}
{r^2\over 6 u} \left({f'^2 \over f^2} - {f'' \over f}
+{\epsilon \over r} {u' \over u}
+{1\over 2}{f' u' \over f u}
-{3\over r}{f' \over f}
-{4 \epsilon \over r^2}
- {6 m^2 \over r^2} u 
\right)
\nonumber
}}\\

The $\Theta$ term is written as the following sum\chg{:}
\ea{\Theta = P_1 + P_2 + P_3} 
where
\ea{\label{P1}
P_1 & = &  
{1\over 4r^2} \bigg\{2\mathcal{Z}_1+\epsilon r^{2\epsilon-2}\left[a_1r-2\epsilon \left({a_2-a_3\over 2}\right)-(2 \epsilon+1)a_3\right]\mathcal{Z}_3 
- r^{4 \epsilon-2}
\left[
\left({a_1r\over 2}-2\left({a_2-a_3\over 2}\right)\epsilon \right)\left(2 \epsilon \sigma+r\sigma'\right) \right.
\nonumber \\&&
\left.-a_3\left(\epsilon (2 \epsilon+1)\sigma-{r^2\sigma''\over 2}\right)
\right]\mathcal{Z}_5 
- \left({a_2-a_3\over 2}\right){r^{4 \epsilon}\over 2}
\left(2\epsilon\sigma r^{\epsilon -1} +r^\epsilon \sigma'\right)^2\mathcal{Z}_7 \bigg\}.
}
\ea{\label{P2}
{P}_2 &=& 
-{2 \over r^{2\epsilon} f} \mathcal{Z}_{-1}
+{a_1 \over 6f} \bigg\{\left({4+2\epsilon \over r}-2{f' \over f}\right)\mathcal{Z}_1 -r^{2\epsilon-1}(2\epsilon\sigma +r \sigma')\mathcal{Z}_3\bigg\} 
- {a_2 -a_3\over 60f}\bigg\{\left({4(3\epsilon^2+4\epsilon+8) \over r^2}-{8(\epsilon+4) \over r}{f' \over f} + 8{f'^2 \over f^2}\right)\mathcal{Z}_1 
\nonumber \\& & 
-4r^{2\epsilon-1}\left({(3\epsilon+2) \over r}-{f' \over f}\right)(2\epsilon\sigma +r \sigma')\mathcal{Z}_3 + 3 r^{4\epsilon-2}(2\epsilon\sigma +r \sigma')^2\mathcal{Z}_5\bigg\} 
- {a_3 \over 6f}\bigg\{\left({12+2\epsilon(2\epsilon+1) \over r^2}-{8 \over r}{f' \over f}+2{f'' \over f}\right)\mathcal{Z}_1 
\nonumber \\& & 
- r^{2\epsilon-2}(2\epsilon(2\epsilon+1)\sigma-r^2\sigma'')\mathcal{Z}_3\bigg\}
}
where we have introduced the \chg{Epstein-Hurwitz} generalized $\zeta$ \chg{function}
\eq{
\mathcal{Z}_q \equiv \sum^{\infty}_{n=1} \left(\omega^2_n+r^{2\epsilon}\sigma(r)\right)^{-q/2}.
\label{Zdef}
}
{For $q\le1$\chg{,} the sum~(\ref{Zdef}) is divergent and is understood as analytically continued (regularized) via its \chg{Chowla-Selberg}/Bessel representation, which also renders it rapidly convergent for numerical evaluation; for $q>1$\chg{,} it converges directly.}
For $P_3$\chg{,} we have
\ea{
P_3  
&=&
P_{31} + P_{32}^{(a)} +
P_{32}^{(b)}
}
where
\ea{
P_{31} &=& {2i\over r^{\epsilon+2}}\sum^{\infty}_{n=1}\int^{\infty}_0{d\tau \over e^{2\pi \tau}-1}
\left[
\left({i\tau+1/2 \over \tilde{W}_n(i\tau)}-{-i\tau+1/2 \over \tilde{W}_n(-i\tau)}\right) 
- i \tau \left(
\left. {2 \over \tilde W_{n}(0)} - \left({1\over \tilde W_{nl}^2} {d \tilde W_{nl} \over d l}\right) \right |_{l\to 0}\right)
\right]
}
\ea{
P_{32}^{(a)} = -{1\over 3} P_1
}
\ea{
P_{32}^{(b)} = {1\over 12 r^{2+\epsilon}} \sum_{n=1}^\infty {1\over W_{nl}^2}{d W_{nl} \over d l} \Big |_{l\to 0}
}
All resulting contributions to~(\ref{renormalizedphisquare}) from $\Delta$ and $\Theta$ are finite and regular.

\end{widetext}

\end{document}